\documentclass[12pt]{iopart}
\usepackage{iopams}
\usepackage{graphicx}
\usepackage{epsfig}
\usepackage{ulem}
\usepackage{cancel}
\usepackage{bm}
\usepackage[flushleft]{threeparttable}
\usepackage[usenames, dvipsnames]{color}
\newcommand{\beq}{\begin{eqnarray}}
\newcommand{\eeq}{\end{eqnarray}}  
\begin{document}

\title[]{Carrier-induced Antisymmetric-symmetric Tendencies of Spin Stiffness in Zigzag Graphene Nanoribbons}

\author{Teguh Budi Prayitno$^1$, Fumiyuki Ishii$^2$,}
 
\address{$^1$Department of Physics, Faculty of Mathematics and Natural Science, Universitas Negeri Jakarta, Kampus A Jl. Rawamangun Muka, Jakarta Timur 13220, Indonesia}
\address{$^2$Nanomaterials Research Institute, Kanazawa University, Kanazawa 920-1192, Japan}
\ead{$^1$teguh-budi@unj.ac.id}
\vspace{10pt}

\begin{abstract}
The generalized Bloch theorem was applied to calculate the spin stiffness and to consider its tendencies when introducing the doping in zigzag graphene nanoribbons. To reach the intentions, two different flat spin spiral formations were constructed by fixing the ferromagnetic and antiferromagnetic spin arrangements at the two different edges by applying a constraint scheme method. A spin stiffness was then calculated by means of a quadratic order function, which maps the total energy difference in the self-consistent calculations onto the Heisenberg Hamiltonian. We found a very high spin stiffness, as predicted previously by the supercell calculation. We also showed that the antisymmetric-symmetric tendencies of spin stiffness are induced by the hole-electron doping. The dependence of ribbon widths of zigzag graphene nanoribbon on the spin stiffness was also provided with similar tendencies when the doping is taken into account.         
\end{abstract}

\noindent{\it Keywords}: spin-wave excitations, spin stiffness, graphene nanoribbon
%\maketitle

\section{Introduction}
The exploration of graphene nanoribbons (GNR) is of interest for the spintronics applications. Present studies confirmed that both zigzag graphene nanoribbons (ZGNR) and armchair graphene nanoribbons (AGNR) are the feasible candidates for the spintronics application by applying the electric field \cite{Rudberg, Kan, Kumar1, Kumar2}, fixing the angle between two different magnetic moment of C atoms at the different edges \cite{Sawada1}, or introducing the doping \cite{Yan, Sawada2, Soriano, Sawada3, Sawada4}. The rare property that has not been fully elucidated is the spin-wave excitations in the GNR. Edwards and Katsnelson predicted that the $sp$ electron systems should have higher spin stiffness than that of 3$d$ transition metals \cite{Edwards}. They analyzed by comparing the dominant contribution between the Stoner excitations and the spin-wave excitations. Previous studies considered the spin-wave excitations using the supercell calculation \cite{Yazyev} or the Hubbard model approach \cite{You, Rhim, Culchac} in the zigzag graphene nanoribbons (ZGNR).    

The aim of this paper is to calculate the spin stiffness and to investigate the tendencies of spin stiffness with respect to the hole-electron doping in the ZGNR by using the generalized Bloch theorem (GBT) within the linear combination of pseudo-atomic orbitals (LCPAO) method. For calculating the spin stiffness, we are given a choice to involve the Heisenberg model or not. In fact, some authors found the reliable spin stiffness in some magnetic systems without using the Heisenberg model \cite{Muiz, Nakamura}, i.e., the spin stiffness is achieved by fitting the total energy difference excluding the magnetic moment. However, for the most cases, such as in 3$d$ transition metals, the Heisenberg model should be used to get the reliable results \cite{Rosengaard, Padja, Shall, Teguh1}. For this situation, Yazyev and Katsnelson proved that the Heisenberg model should be included \cite{Yazyev}. When they ignored the Heisenberg model, they found the spin stiffness of 320 meV{\AA}$^{2}$. Based on Ref. \cite{Edwards}, this value seems less reliable if comparing to the spin stiffness in 3$d$ transition metals. For the comparison, we take fcc nickel with the magnetic moment per atom about 0.6 $\mu_{B}$. The magnetic moment of carbon atom at the edge is about 0.28 $\mu_{B}$, two times smaller than that of fcc nickel. The DFT calculations showed that the spin stiffness of fcc nickel is larger than 700 meV{\AA}$^{2}$ \cite{Rosengaard, Padja, Shall, Teguh1}, so the spin stiffness of ZGNR should be much larger than that of fcc nickel. Yazyev and Katsnelson then obtained a high value of spin stiffness of 2100 meV{\AA}$^{2}$ after considering the Heisenberg model \cite{Yazyev}, which is a more reliable value. The important point is that this high value is related to the allowable length of spin correlation at the low temperature, at which the ZGNR-based devices can work. This means that controlling the spin stiffness, in this case, has a vital role to develop ZGNR-based spintronics applications.   

In this paper, we include the Heisenberg model to calculate the spin stiffness based on the above reason. To employ the GBT, we construct two flat spiral configurations with the constraint scheme method to fix the direction of the magnetic moments of C atoms at two different edges. The frozen magnon method is applied to calculate the magnon energy of the system without considering the spin-orbit interaction. We also show that the magnitude of the magnetic moment is drastically reduced for the high spiral vector. Therefore, the calculation of spin stiffness can only be performed in a set of the lower spiral vectors than that of 3$d$ transition metals, as carried out in our previous paper \cite{Teguh}. We obtain the high spin stiffness for the ferromagnetic and antiferromagnetic spin arrangements at two different edges and find that the tendencies of spin stiffness become different when introducing the doping. The ferromagnetic configuration achieves the antisymmetric tendency while the antiferromagnetic one yields the symmetric tendency.

Note that the calculated spin stiffness via DFT calculation strongly needs verification from the experiment. In the case of the low magnetic moment, such as fcc nickel and ZGNR, the Stoner excitations cannot be neglected. In 3$d$ transition metals, the calculated spin stiffnesses are in good agreement with experiments only in bcc iron and fcc cobalt, whereas the overestimated spin stiffnesses occur in fcc nickel. This situation may be hold for the ZGNR case \cite{Yazyev}. This means that the calculated spin stiffness of ZGNR via DFT calculation could be either underestimated or overestimated with the available experiment. In this paper, the rest of the discussions is organized as follows. The model of ZGNR and the mathematical expression to calculate the spin stiffness will be discussed thoroughly in section 2. We also give the available considerations to simplify the calculation based on the distance of two magnetic carbon atoms at the edges. The calculated spin stiffness and its tendencies when the doping is taken into account will be given in section 3. We also vary the ribbon width of ZGNR to see the tendencies of spin stiffness. We show that the spin stiffness is proportional to the ribbon width of ZGNR. We summarize our discussion based on the results in section 4.  
   
\section{Model of Ferromagnetic and Antiferromagnetic Zigzag Graphene Nanoribbons}
\begin{figure}[h!]
\vspace{-2mm}
\quad\quad\includegraphics[scale=0.6, width =!, height =!]{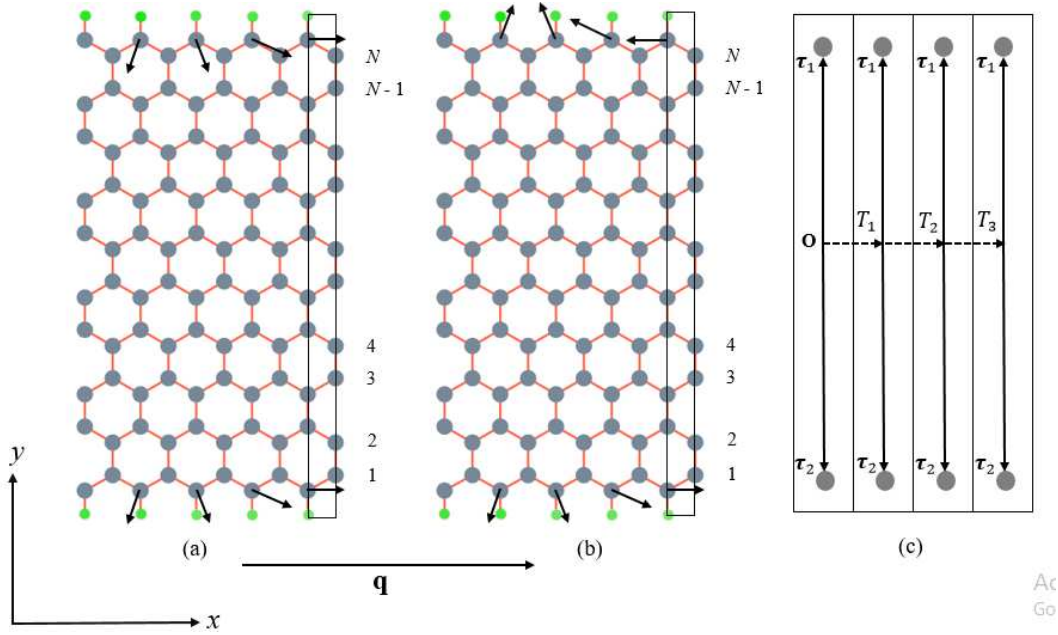}
\vspace{2mm}
\caption{\label{model}(Color online) Flat spiral configurations of ferromagnetic (a) and antiferromagnetic (b) with the ribbon width $N$, and the illustration of the lattice periodicity, as defined by the lattice vectors (c). Here, the gray and green balls represent the C and H atoms, respectively. We draw the black rectangle to represent the unit cell.} 
\end{figure}

To consider the spin-wave excitations using the GBT, the magnetic moment of atoms should be rotated under the direction of a spiral vector $\mathbf{q}$ to establish a spin spiral configuration \cite{Sandratskii}, which is given by 
\beq
	\mathbf{M}_{i}(t)&=&M_{i}(\cos[\varphi^{0}+\mathbf{q}\cdot \mathbf{R}_{i}+\omega_{\mathbf{q}} t]\sin\theta_{i}+\sin[\varphi^{0}+\mathbf{q}\cdot \mathbf{R}_{i}+\omega_{\mathbf{q}} t]\sin\theta_{i}\nonumber\\
		& &+\cos\theta_{i}).\label{moment}
	\eeq
This magnetic moment is characterized by two angles, i.e., the cone angle $\theta$ and the azimuthal angle $\varphi^{\mathbf{q}}$. To consider the spin-wave excitations, the cone angle $\theta$ remains unchanged while the azimuthal angle $\varphi^{\mathbf{q}}$ rotates in the direction of spiral vector $\mathbf{q}$ defined as $\varphi_{i}^{\mathbf{q}}(t)=\mathbf{q}\cdot \mathbf{R}_{i}+\omega_{\mathbf{q}} t$, where ${R}_{i}$ is the lattice vector and $\omega_{\mathbf{q}}$ is the frequency of magnon. The first-principles calculation in this paper was carried out by using the OpenMX code \cite{Openmx}, which applies the LCPAO method and the norm-conserving pseudopotentials \cite{Troullier}. The implementation of GBT in the OpenMX code was conducted by writing the Bloch wavefunctions on the atomic site $\tau_{i}$ in terms of an LCPAO as \cite{Teguh} 
 \beq
\left<\mathbf{r}|\psi_{\nu\mathbf{k}}\right>=\psi_{\nu\mathbf{k}}\left(\mathbf{r}\right)&=&\frac{1}{\sqrt{N}}\left[\sum_{n}^{N}e^{i(\mathbf{k}-\mathbf{q}/2)\cdot\mathbf{R}_{n}}\sum_{i\alpha}C_{\nu\mathbf{k},i\alpha}^{\uparrow}\phi_{i\alpha}\left(\mathrm{\mathbf{r}-\tau_{i}-\mathbf{R}_{n}}\right)
\left(
\begin{array}{cc}
1\\
0\end{array} 
\right)\right.\nonumber\\
& &\left.+\sum_{n}^{N}e^{i(\mathbf{k}+\mathbf{q}/2)\cdot\mathbf{R}_{n}}\sum_{i\alpha}C_{\nu\mathbf{k},i\alpha}^{\downarrow}\phi_{i\alpha}\left(\mathrm{\mathbf{r}-\tau_{i}-\mathbf{R}_{n}}\right)\left(
\begin{array}{cc}
0\\
1\end{array}
\right)\right],\label{lcpao}
\eeq  
where the pseudo-atomic orbital (PAO) $\phi_{i\alpha}$ is constructed by the confinement scheme method \cite{Ozaki1, Ozaki2}. For applying the GBT, we first build the model of ferromagnetic and antiferromagnetic configurations at two different edges of ZGNR, as shown in Figs. \ref{model}(a) and \ref{model}(b), where the structural optimization was carried out by using the nonmagnetic state. The flat spiral configuration will be then generated by fixing the cone angle $\theta=\pi/2$ with the applied penalty functional in the constraint scheme method. The penalty functional works if the direction of the magnetic moment deviates away from the initial $\theta$ in the self-consistent calculation. This functional will force the deviated direction of the magnetic moment back to the initial condition, for the detailed explanations on the constraint scheme method one can see Refs. \cite{Gebauer, Kurz, Cuadrado}. So, the total energy and the magnetic moment will then be evaluated self-consistently with a fixed $\theta$, where $\varphi$ rotates in the $\mathbf{q}$ direction.

We assume that the total energy of the system can be written in terms of Heisenberg Hamiltonian as
\beq
E&=&E_{0} -\frac{1}{2N} \sum_{i\neq j}J_{ij}\mathbf{M}_{i}\cdot\mathbf{M}_{j}\nonumber\\
     &=& E_{0}-\frac{1}{2N} \sum_{i\neq j}J_{ij}M_{i}M_{j}\left\{\cos\left[\mathbf{q}\cdot \left(\mathbf{R}_{i}-\mathbf{R}_{j}\right)\right]\sin\theta_{i}\sin\theta_{j}\right.\nonumber\\
		& &\left.+\cos\theta_{i}\cos\theta_{j}\right\},
		\label{Heisenberg} 
\eeq
where $N$ defines the number of unit cells and $E_{0}$ is the ground state energy. To consider the periodicity in our system, we rewrite the lattice vector ${R}_{i}$ with the double index notation ${R}_{m+\mu}=T_{m}+\tau_{\mu}$ as the linear combination of the translation vector $T_{m}$ and atomic site $\tau_{\mu}$ \cite{Essenberg}, see Fig. \ref{model}(c).  Since the frozen magnon method works in the reciprocal space, it is convenient to transform the exchange coupling constant $J_{ij}$ in the real space to the reciprocal space by means of the Fourier transform  
\beq
J_{\mathbf{q}}^{\mu\nu}=-\sum_{m}(1-\delta^{0m}\delta^{\mu\nu})J^{\mu(\nu+m)}e^{-i\mathbf{q}(\mathbf{R}_{\nu+m}-\mathbf{R}_{\mu})}. \label{Fourier}
\eeq
From Eq. (\ref{Fourier}), it can be derived the following symmetry properties of the exchange coupling constant in the reciprocal space 
\beq
J_{\mathbf{q}}^{\mu\nu}=J_{-\mathbf{q}}^{\nu\mu}, \quad \textrm{Re}[J_{\mathbf{q}}^{\mu\nu}]=\textrm{Re}[J_{\mathbf{q}}^{\nu\mu}], \quad \textrm{Im}[J_{\mathbf{q}}^{\mu\nu}]=-\textrm{Im}[J_{\mathbf{q}}^{\nu\mu}]. \label{symmetry}
\eeq

Applying the small deviation of $\theta$ ($\theta<<$), the total energy in Eq. (\ref{Heisenberg}) then becomes \cite{Halilov}
 \beq
E=E_{0}+\frac{1}{2} \sum_{\mu\nu}M_{\mu}M_{\nu}\left\{(1-\theta^{2}_{\mu}/2-\theta^{2}_{\nu}/2)J_{0}^{\mu\nu}+\theta_{\mu}\theta_{\nu}\textrm{Re}[J_{\mathbf{q}}^{\mu\nu}]\right\}.\label{Heisen_small} 
\eeq
Since we deal with the small $\theta_{\mu}$, we can derive the exchange coupling constant in terms of second derivative of the total energy with respect to $\theta_{\mu}$  
\beq
\textrm{Re}[\tilde{J}_{\mathbf{q}}^{\mu\nu}]=\left[\frac{1}{M_{\mu}M_{\nu}}\frac{\partial^{2}E}{\partial\theta_{\mu}\partial\theta_{\nu}}\right]_{\theta=0},\label{diff}
\eeq 
where we define
\beq
\textrm{Re}[\tilde{J}_{\mathbf{q}}^{\mu\nu}]=J_{\mathbf{q}}^{\mu\nu}-\delta^{\mu\nu}\sum_{\alpha}\frac{M_{\alpha}}{M_{\mu}}J_{0}^{\alpha\mu}.\label{def_J}
\eeq 

The next step is to formulate the eigenvalue equation to calculate the magnon energy. We consider the quantum version of the Heisenberg motion for the operator of magnetic moments. 
\beq
i\hbar\frac{d\hat{\textit{\textbf{M}}}_{i}}{dt}=\left[\hat{\textit{\textbf{M}}}_{i},\hat{H}\right],\label{dynmom}
\eeq 
with the commutation relations
\beq
\left[\hat{\textit{\textbf{M}}}_{i}^{\alpha},\hat{\textit{\textbf{M}}}_{j}^{\beta}\right]=i\mu_{B}\delta_{ij}\epsilon_{\alpha\beta\gamma}\hat{\textit{\textbf{M}}}_{i}^{\gamma}.\label{commut}
\eeq   
 In this case, the Greek and Roman indexes refer to the Cartesian coordinates and the lattice sites. Note that the commutation relation in Eq. (\ref{commut}) is referred to the definition of the operator of magnetic moment at the lattice site $i$
\beq
\hat{\textit{\textbf{M}}}_{i}(t)=\mu_{B}\int_{V_{i}}d^{3}\mathbf{r}\psi^{\dagger}(\mathbf{r},t)\hat{\bm{\sigma}}\psi(\mathbf{r},t),
\eeq   
where $V_{i}$ is the crystal volume and $\hat{\bm{\sigma}}$ denotes the operator of Pauli matrices. Extracting Eq. (\ref{dynmom}) and Eq. (\ref{commut}) in terms of the components of operator of magnetic moment, we obtain the following equations
 \beq
\hbar\sin\theta_{i}\frac{d\varphi_{i}^{\mathbf{q}}}{dt}=\mu_{B}\sum_{j(\neq i)}J_{ij}M_{j}[\sin\theta_{i}\cos\theta_{j}-\cos\theta_{i}\sin\theta_{j}\cos(\varphi_{i}^{\mathbf{q}}-\varphi_{j}^{\mathbf{q}})],\label{one}\\
\hbar\frac{d\theta_{i}}{dt}=-\mu_{B}\sum_{j(\neq i)}J_{ij}M_{j}\sin\theta_{i}\sin(\varphi_{i}^{\mathbf{q}}-\varphi_{j}^{\mathbf{q}}),\label{two}
\eeq  
as well as the time-independent of the magnitude of magnetic moment $dM_{i}/dt=0$. By applying the small $\theta$ and the Fourier transform to $J_{ij}$ in Eq. (\ref{Fourier}), those equations above yield the eigenvalue problem  
\beq
\sqrt{M_{\mu}}\theta_{\mu}\hbar\omega_{\mathbf{q}}=\mu_{B}\sum_{\nu}\sqrt{M_{\mu}M_{\nu}}\textrm{Re}[\tilde{J}_{\mathbf{q}}^{\mu\nu}]\sqrt{M_{\nu}}\theta_{\nu}.\label{eigen_value}
\eeq 
We can deduce that Eq. (\ref{eigen_value}) corresponds to the eigenvector $\sqrt{M_{\mu}}\theta_{\mu}$ and the eigenvalue $\hbar\omega_{\mathbf{q}}$ obtained by diagonalizing the matrix $\mu_{B}\sqrt{M_{\mu}M_{\nu}}\textrm{Re}[\tilde{J}_{\mathbf{q}}^{\mu\nu}]$. Thus, to obtain the magnon energy, we have to solve the following eigenvalue problem 
\beq
\textrm{det}(\delta_{\mu\nu}\hbar\omega_{\mathbf{q}}-\mu_{B}\sqrt{M_{\mu}M_{\nu}}\textrm{Re}[\tilde{J}_{\mathbf{q}}^{\mu\nu}])=0.\label{eigen_final}
\eeq

As directly seen in Figs. \ref{model}(a) and \ref{model}(b), the two models of ZGNR contain two magnetic carbon atoms (two sublattices) in the unit cell with the same magnetic moment $M_{1}=M_{2}=M$. In this case, the ferromagnetic configuration indicates the initial angles $\theta_{1}=\theta_{2}=\pi/2$ and $\varphi_{1}^{0}=\varphi_{2}^{0}=0$, while the antiferromagnetic one indicates the initial angles $\theta_{1}=\theta_{2}=\pi/2$ and $\varphi_{1}^{0}=0$, $\varphi_{2}^{0}=\pi$. Consequently, the eigenvalues of the magnon energy attained from Eq. (\ref{eigen_final}) are given by
   \beq
	\hbar\omega_{\mathbf{q}}&=&\frac{1}{2}\mu_{B}M(\textrm{Re}[\tilde{J}_{\mathbf{q}}]^{11}+\textrm{Re}[\tilde{J}_{\mathbf{q}}^{22}])\nonumber\\
	& &\pm \frac{1}{2}\mu_{B}M\sqrt{4(\textrm{Re}[\tilde{J}_{\mathbf{q}}^{12}])^{2}+(\textrm{Re}[\tilde{J}_{\mathbf{q}}]^{11}-\textrm{Re}\tilde{J}_{\mathbf{q}}^{22}])^{2}}, \label{eigen}
	\eeq
where $\textrm{Re}[\tilde{J}_{\mathbf{q}}^{11}]$ is the exchange interaction between C atoms in one edge, $\textrm{Re}[\tilde{J}_{\mathbf{q}}^{22}]$ is the exchange interaction between C atoms in the other edge, and $\textrm{Re}[\tilde{J}_{\mathbf{q}}^{12}]$ is the exchange interaction between C atom in one edge and C atom in the other edge. Since the distance of two magnetic carbon atoms at the different edges in the unit cell is much larger than the length of periodic unit cell for the two configurations in Figs. \ref{model}(a) and \ref{model}(b), we are allowable to approximate Eq. (\ref{eigen}) to obtain two magnon energies
		\beq
	\hbar\omega_{\mathbf{q}1}=\mu_{B}M\textrm{Re}[\tilde{J}_{\mathbf{q}}^{11}], \qquad  \hbar\omega_{\mathbf{q}2}=\mu_{B}M\textrm{Re}[\tilde{J}^{22}_{\mathbf{q}2}]. \label{eigen_approximate}
	\eeq
Since two magnetic C atoms, in this case, are equivalent, $\textrm{Re}[\tilde{J}_{\mathbf{q}}^{11}]$ and $\textrm{Re}[\tilde{J}_{\mathbf{q}}^{22}]$ become equivalent, too. Thus, we can reduce the computation by performing one calculation to obtain the magnon energy, as carried out by Essenberg $et$ $al$ \cite{Essenberg}. In this case, we make a different approach with theirs for obtaining the final expression of magnon energy. Realizing $\theta_{1}=\theta_{2}=\theta$, Eq. (\ref{Heisen_small}) then becomes
\beq 
E=E_{0}+M^{2}(1-\theta^{2})J^{11}_{0}+M^{2}\theta^{2}\textrm{Re}[J_{\mathbf{q}}^{11}].\label{eigen_reduce} 
\eeq   
We then differentiate Eq. (\ref{eigen_reduce}) with respect to $\theta$ to obtain
\beq 
\textrm{Re}[\tilde{J}_{\mathbf{q}}^{11}]=\frac{1}{2M^{2}}\frac{\partial^{2}E}{\partial\theta^{2}}.\label{J} 
\eeq   
By replacing the left-hand side using Eq. (\ref{eigen_approximate}) and replacing the second derivative of total energy in terms of the total energy difference at the right-hand side, Eq. (\ref{J}) is cast into  
\beq 
\hbar\omega_{\mathbf{q}}=\frac{\mu_{B}}{M}\frac{\Delta E(\mathbf{q},\theta)}{\sin^{2}\theta},\label{magnonf} 
\eeq 
where $\Delta E(\mathbf{q},\theta)=E(\mathbf{q},\theta)-E(\mathbf{0},\theta)$ is the total energy difference. Note that, one can get Eq. (\ref{J}) by employing the L'H$\hat{\textrm{o}}$pital's theorem at the right-hand side of Eq. (\ref{magnonf}). Although the derivation employs the small $\theta$, the similar magnon dispersion relation formulated in Eq. (\ref{magnonf}) can also hold for the larger $\theta$, for example see Refs. \cite{Enkovaara, Jakobson, Lezaic}.

\section{Results}
\begin{figure}[h!]
\vspace{-4mm}
\quad\quad\includegraphics[scale=0.5, width =!, height =!]{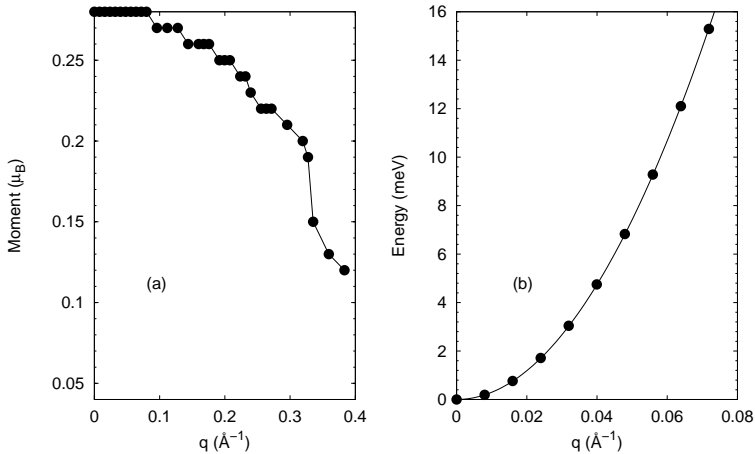}
\vspace{-58mm}
\caption{\label{mom-stif} The reduction of the magnetic moment as $\mathbf{q}$ increases (a) and the magnon energy curve close to $\mathbf{q}=0$ (b). These figures are achieved from the ferromagnetic configuration. The antiferromagnetic one has the same tendency, too.} 
\end{figure}

The computation was performed by a $90 \times 1 \times 1$ $k$ point sampling in the Brillouin zone with the generalized gradient approximation (GGA) for the exchange-correlation potential \cite{Perdew}. For the basis sets, we used two valence orbitals $s$ and two valence orbitals $p$ for C atoms with the cutoff radius of 4.0 a.u. (atomic unit). Meanwhile, we used two valence orbitals $s$ and one valence orbital $p$ for H atoms with the cutoff radius of 6.0 a.u. We also used the cutoff energy of 150 Ryd to ensure the convergence of self-consistent calculations. In addition, we set the experimental lattice constant of graphite of 2.46 {\AA} for the length of the periodic unit cell ($x$ axis) and fixed the length of the nonperiodic cell larger than 25 {\AA} to create a sufficient vacuum condition. For the first discussion, we focused on the ribbon width $N=10$. Since the self-consistent calculations using the frozen magnon approach must be carried out for each $\mathbf{q}$, at which the magnetic moment is nearly constant, we set the nonzero $\mathbf{q}$ very close to $\mathbf{q}=0$. This is due to the drastic reduction of the magnetic moment as $\mathbf{q}$ increases, as shown in Fig. \ref{mom-stif}(a).

After calculating the magnon energy for each $\mathbf{q}$ using Eq. (\ref{magnonf}), we fit the data by using the quadratic dispersion $E=D\mathbf{q}^{2}$ to obtain the spin stiffness $D$. Figure \ref{mom-stif}(b) shows the spectrum of magnon energy of 10-ZGNR for the ferromagnetic configuration for the nondoped case. Our calculations yield $D=2966$ meV{\AA}$^{2}$ for the ferromagnetic configuration and $D=3583$ meV{\AA}$^{2}$ for the antiferromagnetic one for the nondoped case. Note that the antiferromagnetic configuration is more stable than the ferromagnetic one. We see that the spin stiffness of the antiferromagnetic configuration is larger than the ferromagnetic one due to the higher magnon energy of the antiferromagnetic one. This means that the spin stiffness of the most stable configuration in 10-ZGNR should have the highest spin stiffness. We also find that the magnetic moment of the antiferromagnetic configuration is always higher than that of the ferromagnetic one. This means that there is a close relationship between the spin stiffness and the magnetic moment. We will see more clearly this relationship when introducing the doping or varying the ribbon width in the next discussion.
\begin{figure}[h!]
\vspace{2mm}
\quad\quad\includegraphics[scale=0.5, width =!, height =!]{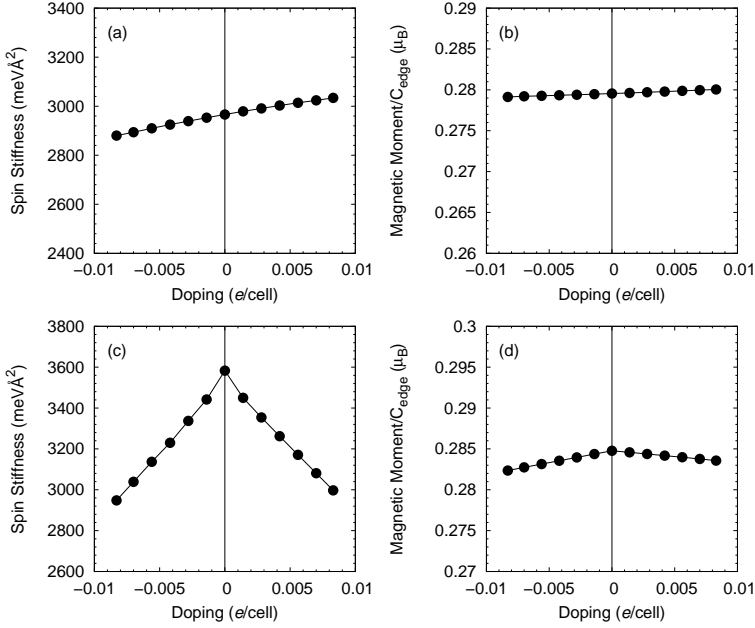}
\vspace{-34mm}
\caption{\label{stiff-doped} The tendencies of spin stiffness and magnetic moment for 10-ZGNR when the doping is considered. Figures (a) and (c) represent the spin stiffness of FM state and AFM state, while figures (b) and (d) denote the magnetic moment per C$_{\textrm{edge}}$ atom of FM state and AFM state. It is shown that both the spin stiffness and the magnetic moment follow the same tendency, i.e., the FM and AFM states show the antisymmetric and symmetric tendencies, respectively.}
\end{figure}

To introduce the hole-electron doping as the uniform background charge, we employ the Fermi level shift method to keep the system neutral. We find the different tendencies for the ferromagnetic and antiferromagnetic configurations for the hole-electron doping. As shown in Fig. \ref{stiff-doped}, for the ferromagnetic configuration, as the hole (electron) doping increases, the spin stiffness increases (reduces). This situation is similar to our last paper \cite{Teguh}, which used the fourth-order fit to obtain the spin stiffness. The difference between the quadratic dispersion and the fourth-order dispersion lies in the fitting error. The fitting error of quadratic dispersion is usually higher than that of the fourth-order dispersion. However, the quadratic dispersion, in this case, is sufficient to produce the curve of magnon energy, as shown in Fig. \ref{mom-stif}(b). For the antiferromagnetic case, on the contrary, the situation is very different. As the hole (electron) doping increases, the spin stiffness reduces (reduces). Therefore, by introducing the hole-electron doping, the ferromagnetic and antiferromagnetic configurations create the antisymmetric and symmetric tendencies, respectively. These tendencies also hold for the magnetic moments, which almost follow the tendencies of spin stiffness. Note that the magnetism in ZGNR disappears if the introduced doping is sufficiently high.   
\begin{figure}[h!]
\vspace{2mm}
\quad\quad\includegraphics[scale=0.5, width =!, height =!]{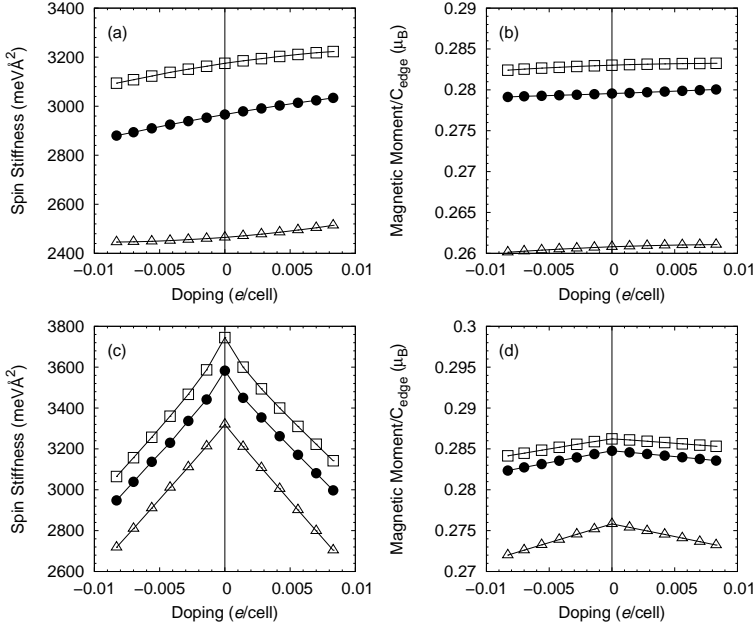}
\vspace{-34mm}
\caption{\label{N-stiff-doped} The tendencies of spin stiffness and magnetic moments for $N$-ZGNR, where $N=6, 10, 12$ are represented by the triangles, filled circles, and boxes, respectively. Figures (a) and (c) represent the spin stiffness of FM state and AFM state, while figures (b) and (d) denote the magnetic moment per C$_{\textrm{edge}}$ atom of FM state and AFM state. It is shown that the antisymmetric and symmetric tendencies for the FM and the AFM states also hold for the other ribbon widths. In addition, the ribbon width determines the spin stiffness and the magnetic moment.}
\end{figure}

The last discussion is to see the dependence of spin stiffness on the ribbon width. As shown in Fig. \ref{N-stiff-doped}, we see that the spin stiffness is proportional to the ribbon width. As the ribbon width increases, the spin stiffness increases, see table \ref{N-stiff}. This means that the required energy becomes higher when the ribbon width increases to make the spin-wave excitations. We also show that all the tendencies for 10-ZGNR also hold for the other ribbon widths when introducing the hole-electron doping. The same tendencies between the spin stiffness and the magnetic moment are due to the mapping of the energy of Heisenberg Hamiltonian onto the total energy difference in the self-consistent calculation. When the exchange constant $J_{ij}$ remains unchanged, the energy of Heisenberg Hamiltonian is proportional to the magnetic moment. At the same situation, the spin stiffness proportionally depends on the total energy difference in the self-consistent calculation. Therefore, the spin stiffness and the magnetic moment should have the same tendency. 
\begin{table}
\vspace{3 mm}
\caption{The spin stiffness dependence of the ribbon width for the nondoped case.}   % title of Table
\centering % used for centering table
\begin{tabular}{c c c} % centered columns (6 columns)
\hline % inserts single horizontal line
Width& $D_{\textrm{FM}}$ (meV{\AA}$^{2}$)&$D_{\textrm{AFM}}$ (meV{\AA}$^{2}$)\\ % inserting body of the table
\hline %inserts single line
6& 2465&3320\\ % [1ex] adds vertical space
10& 2966&3583\\ % [1ex] adds vertical space
12& 3175&3745\\ % [1ex] adds vertical space
\hline %inserts single line
\end{tabular}
\vspace{0.01cm}
\label{N-stiff} % is used to refer to this table in the text
\vspace{-6 mm}
\end{table}

\section{Conclusions} 
We obtain the very high spin stiffness for $N$-ZGNR compared to 3$d$ transition metals, as predicted previously by Edwards and Katsnelson \cite{Edwards}. Compared to Yazyev and Katsnelson's result by the supercell calculation \cite{Yazyev}, our calculated spin stiffnesses of both the ferromagnetic and antiferromagnetic configurations are higher but still in the same order. We also show that the antiferromagnetic configuration, which is more stable than the ferromagnetic one, has a higher spin stiffness than that of the ferromagnetic one. This implies that the most stable configuration in $N$-ZGNR should have the highest spin stiffness. It is also shown that the spin stiffness increases as the ribbon width increases. It suggests that the distance of two magnetic carbon atoms at the different edges in the unit cell determines the high or low spin stiffness.

When the doping is taken into account, we find the different tendencies of spin stiffness. For the ferromagnetic (antiferromagnetic) configuration, as the hole doping increases, the spin stiffness increases (reduces). Contrarily, the spin stiffness reduces (reduces) as the electron doping increases for the ferromagnetic (antiferromagnetic) configuration. It seems that introducing the doping can control the spin stiffness of $N$-ZGNR.     
%-------------------------------------------------------------------------------
\section*{Acknowledgments}
The computations were partly performed using ISSP supercomputers at the University of Tokyo, while the rest of the computations was performed at the Universitas Negeri Jakarta. This work was supported by Japan Society for the Promotion of Science (JSPS) Grants-in-Aid for Scientific Research on Innovative Area, “Nano Spin Conversion Science” (Grant Nos. 15H01015 and 17H05180). It was also supported by a JSPS Grant-in-Aid for Scientific Research on Innovative Area, “Discrete Geometric Analysis for Material Design” (Grant No. 18H04481). It was partially supported by a JSPS Grant-in-Aid on Scientific Research (Grant No. 16K04875)

\end{document}